\documentstyle[12pt,epsf]{article}

\newcommand{\be}{\begin{equation}}
\newcommand{\ee}{\end{equation}}
\newcommand{\bn}{\begin{eqnarray}}
\newcommand{\en}{\end{eqnarray}}
\newcommand{\bd}{\begin{displaymath}}
\newcommand{\ed}{\end{displaymath}}

\begin{document}

\begin{flushright}
IFT-P.059/97
\\hep-th/9709205
\end{flushright}
\vspace{0.5cm}
\begin{center}
{\Large \bf Aspects of classical and quantum motion \\ on a flux cone}
\footnote{Work supported by FAPESP grant 96/12259-1.} 
\\ 
\vspace{1cm} 
{\large E. S. Moreira Jnr.}\footnote{e-mail: moreira@axp.ift.unesp.br}   
\\ 
\vspace{0.3cm} 
{\em Instituto de F\'{\i}sica  Te\'{o}rica} \\ 
{\em Universidade Estadual Paulista,}   \\
{\em Rua Pamplona, 145}  \\
{\em 01405-900 - S\~{a}o Paulo, S.P., Brazil}  \\
\vspace{0.3cm}
{\large August, 1997}
\end{center}
\vspace{1cm}

\begin{abstract}
Motion of a non-relativistic particle on a cone with a 
magnetic flux running through the cone axis (a ``flux cone'') is
studied. It is expressed as the motion of 
a particle moving on the Euclidean plane under the action of 
a velocity-dependent force. Probability fluid (``quantum flow'')
associated with a particular stationary state 
is studied close to the singularity, demonstrating non trivial 
Aharonov-Bohm effects. For example, it is shown 
that near the singularity quantum flow departs 
from classical flow. In the context of the
hydrodynamical approach to quantum mechanics,
quantum potential due to the conical singularity is 
determined and the way it affects quantum flow is analysed.
It is shown that the winding number of classical orbits 
plays a role in the description of the quantum flow.
Connectivity of the configuration space is also discussed.
\end{abstract}

\section{Introduction}
Usually, when the
quantum description of a system is non trivial, so is the
classical one. However this is not always the case. 
There is a number of examples whose intrinsic quantum effects
do not have a classical analogue. Notable among these is the
Aharonov-Bohm (A-B) effect \cite{aha59}. 
In this set up, the magnetic field vanishes everywhere except inside
a thin flux tube. As there is no Lorentz force, classically particles
are free, and are not affected by the background.
However in the quantum scattering problem
the background leads to a non trivial scattering, 
which is confirmed by experiment \cite{ola85}.
A curious fact about the A-B effect is that
by choosing an appropriate gauge the quantum mechanical 
equations of motion become free equations and the interaction 
of quantum matter with a singular vector
potential becomes encoded in unconventional boundary conditions.

The A-B effect is a consequence of the interaction of quantum matter
with a {\it nearly} trivial affine connection and it is present in every gauge
theory, including gravity.
The analogue of the A-B set up in gravitation is a conical background
\cite{mar59,sta63,dow67}. The geometry is flat everywhere apart from a 
symmetry axis. As in the case of a thin flux tube, the
problem of studying quantum theory in this background amounts 
to solving the usual equations in flat space,
with the non trivial conical geometry manifested only 
in the boundary conditions. Solutions of these equations lead to
{geometrical} Aharonov-Bohm effects. 
It should be remarked  that there is  
an important distinction between  A-B effects in ordinary
gauge theories and those in gravity. In ordinary gauge theories, 
matter does not couple with (non trivial) connections in the classical 
equations of motion but with (vanishing)
field strengths. This is not the case in gravity
where (non trivial) connections are also present in the classical equations
of motion, making the effect appear even in the classical
theory. An interesting example is the 
self-force induced by conical singularities
\cite{lin86,smi90}. Another is the double images effect due to 
cosmic strings, whose external gravitational field may approximately
be described by a conical geometry \cite{vil94}.

Quantum theory on cones has been studied over two decades 
\cite{dow77},
and investigations intensified in the mid-eighties 
\cite{smi90,hel86}
very much due to
the importance in cosmology of cosmic strings.
More recently interest in the subject was renewed 
in the context of quantum mechanics of black holes, 
where one encounters conical singularities
(see e.g. \cite{zer96} and references therein).
Besides these and other practical motivations to study 
quantum theory in conical backgrounds, there is another more 
academic one [12-14].
Namely,
the study of quantum mechanics in a {\it nearly} trivial gravitational 
background may shed some light on the profound problems of 
combining quantum mechanics and general relativity.

In this work classical and quantum effects caused by a conical
sigularity on the motion of a particle are studied.
On various occasions the 
A-B set up is coupled to a conical geometry (flux cone), so that
a detailed comparison of the corresponding A-B effects is possible.
The paper is organized as follows.
A study of the conical geometry is given in section 2,
commenting on points which have been overlooked in the literature.
In particular the Aharonov-Bohm like features of the 
Levi-Civita connection are stressed. 
A regularization method of determining the localized
curvature of a cone is proposed. 
Positive and negative deficit angles
are considered and related with the curvature. 
The relevant coordinate systems are defined. 
In section 3
classical motion on a flux cone is studied. 
The effects of the conical singularity are shown to
be equivalent to the ones due to an angular momentum dependent force.
In section 4
quantization is implemented by the usual substitution principle.
The issue of boundary conditions at the singularity is considered,
and a particular one is chosen, motivated by 
regularization arguments given in \cite{kay91,bou92}.
Corresponding stationary states are obtained
and used to build up a state to probe the 
singularity.
This leads, in section 5, to the study of quantum flow, 
showing new non trivial 
effects due to the conical singularity. 
Such effects are caused by a non
vanishing quantum potential whose features are mentioned.
Connectivity of the configuration space is discussed by taking into
account the  behaviour of quantum flow at the singularity.
The last section is a summary, including possible extensions of this
study. An account on the hydrodynamical approach to quantum mechanics
(whose elements are used in section 5) is given in the appendix.

\section{The cone}
\begin{figure}
\centerline{\epsffile{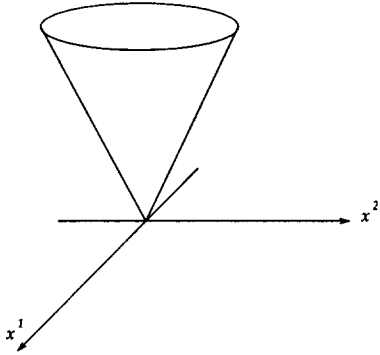}}
\caption{{\small Cone embedded in a 3-dimensional Euclidean space}.} 
\label{econe}
\end{figure}

A cone is a 2-dimensional space with a $\delta$-function curvature
singularity.
The curvature tensor is concentrated at a single point, vanishing 
everywhere else \cite{sok77}. The line element may be written as 
\bn
dl^{2}&=&g_{ij}dx^{i}dx^{j} \nonumber\\
&=&\left[ \delta_{ij}+ \left( \alpha^{-2}-1 \right) 
\frac{x^{i}x^{j}}{r^{2}}
\right] dx^{i}dx^{j},
\label{dlc}
\en
where the coordinate 
$x^{i}$ $ \left( i=1, 2\right) $ runs from $-\infty $ 
to  $\infty $ , $r^{2}:= \delta_{ij} x^{i}x^{j}$ and
$\alpha$ is a positive parameter \cite{des84}. 
Imagining the conical surface embedded in a 3-dimensional Euclidean space,
$\left\{ x^{1}, x^{2}\right\} $ are 
Cartesian coordinates on a plane perpendicular 
to the symmetry axis of the cone (Fig. \ref{econe}).
From (\ref{dlc}) one sees that the conical singularity is located at 
the origin and that when $\alpha=1$ the cone becomes the Euclidean plane.

\begin{figure}
\centerline{\epsffile{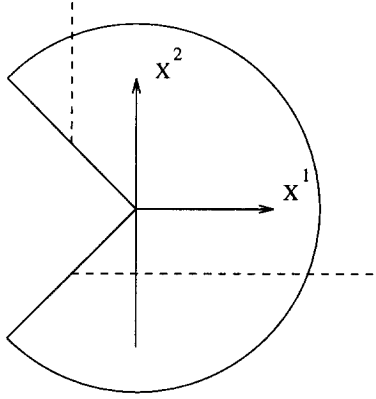}}
\caption{{\small Singular Cartesian frame}.} 
\label{ocone}
\end{figure}

Using polar coordinates 
($x^{1}=r\cos \theta$, 
$ x^{2}=r\sin \theta$), 
the line element (\ref{dlc}) can be rewritten as
$dl^{2}=\alpha^{-2} dr^{2}+r^{2}d\theta^{2}$
where the conical singularity is now hidden 
by the coordinate singularity 
at the origin,
since the polar angle $\theta$ is not defined at $r=0$. 
Note that the coordinates $\left(r,\theta \right)$ 
and $\left(r,\theta +2\pi\right)$
label the same point,
$\left(r,\theta \right)\sim \left(r,\theta +2\pi\right)$.
A further simplification of the line element can be made by rescaling 
$\left\{r,\theta\right\}$ as 
\be
\rho =\alpha^{-1}r \hspace{1.0cm}
\varphi =\alpha \theta, 
\label{npol}
\ee
resulting in
\be
dl^{2}=d\rho^{2}+\rho^{2}d\varphi^{2}, 
\label{dlP}
\ee
which is the line element of the Euclidean plane written in polar coordinates,
showing the flatness of the cone.
However, as a consequence of the rescaling, now the coordinates
$\left(\rho ,\varphi \right)$ 
and 
$\left(\rho ,\varphi +2\pi\alpha \right)$ 
label the same point,
\be
\left(\rho ,\varphi \right) \sim  \left(\rho ,\varphi +2\pi\alpha \right). 
\label{Boun}
\ee
This unusual identification encapsulates the fact that the space is
a cone and not the Euclidean plane. The coordinates
$\left\{ \rho , \varphi \right\}$
are polar coodinates on the surface of the cone,
and they may be  
visualized by cutting the cone along its generating line and opening it flat.
The angle of the missing wedge (extra wedge, when $\alpha >1$)
is the deficit angle of the cone 
${\cal D}= 2\pi \left(1-\alpha \right)$ (Fig. \ref{ocone}).

\begin{figure}
\centerline{\epsffile{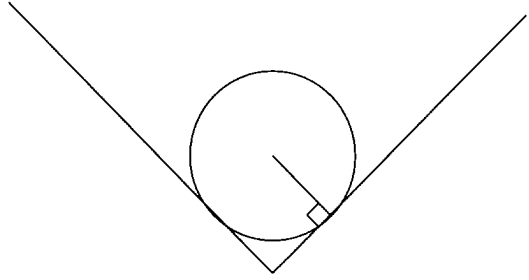}}
\caption{{\small Smoothing the cone}.} 
\label{scone}
\end{figure}

At this point it should be noted that the Euclidean plane and a cone have the 
same topology, differing in their geometry -  the former is globally flat
whereas the latter is not. To see this, one smoothes the cone by
replacing its tip by a tangential spherical 
cap of radius $a$ (Fig. \ref{scone}).
Clearly the resulting surface is simply connected, a fact that does not
change when $a\rightarrow 0$ and the idealized cone is recovered.
As the curvature scalar of the spherical cap is given by $R=-2/a^{2}$,
one is left with a curvature singularity surrounded by a flat surface.
By considering the line element on the sphere, it may be easily shown that
\be
\int d^{2}x\ \sqrt{g}R =-2{\cal D},
\label{ir}
\ee
which demonstrates the delta function character of the conical singularity.
Note that though the cone
is a simply connected background, the configuration space of a quantum 
particle living on it may be non simply connected (see section 5).

It follows from (\ref{Boun}) that the ``Cartesian'' coordinates
defined by 
$X^{1}:=\rho \cos \varphi$ and $ X^{2}:=\rho \sin \varphi$
(Fig. \ref{ocone}) 
are singular if ${\cal D} \not= 0$: in terms of 
$\left\{X^{1}, X^{2}\right\}$ 
the metric tensor is Euclidean everywhere 
except on the rays defining the borders of the wedge, 
where these coordinates are 
discontinuous functions of $\varphi $ 
(In order to obtain the ``Euclidean'' metric tensor 
from (\ref{dlP}) one has to differentiate
a discontinuous function.).
The borders of the wedge can be arbitrarily rotated by redefining 
the interval of length $2\pi\alpha$ over which $\varphi$ ranges.
For example, when $0\leq \varphi <2\pi\alpha$ the borders are at
$\varphi=0 \sim 2\pi\alpha$, and at
$\varphi = -\pi \alpha \sim  \pi\alpha$ when
$-\pi \alpha\leq \varphi <\pi\alpha$,
which is the case illustrated in Fig. \ref{ocone}.

Corresponding to the behaviour of the metric tensor
in terms of $\left\{X^{1}, X^{2}\right\}$, the 
Levi-Civita connection vanishes everywhere except on
the borders of the wedge where it is singular.
It should be stressed that the connection does not 
vanish everywhere around the singularity
in contradiction to what is sometimes claimed.
If that were the case, a vector parallely propagated
on a closed loop around the singularity would 
match itself as the loop is completed \cite{dow67}, 
which clearly does not happen
as long as there is a wedge. The non-vanishing connection
tells the rest of the space that there is a curvature
singularity at the origin, and this non trivial behaviour
is the very one responsible for the geometrical Aharonov-Bohm like
effects that will be seen in this work.

It is worth remarking that the rays where the 
metric tensor is singular come into the
problem only if one insists on using the singular
Cartesian coordinate system. They are a coordinate
singularity.
Note that for $\alpha>1$
more than one
singular Cartesian coordinate system is needed to cover the whole cone.
In this case there is a ``branch cut'' leading 
to a new Cartesian coordinate system
in which there is an  
extra wedge (with corresponding negative
deficit angle).

In the following sections, both frames $\left\{X^{1}, X^{2}\right\}$
and $\left\{x^{1}, x^{2}\right\}$
will be used to study motion on the cone.
It will be clear from the text which frame 
is used in each situation.

\section{Classical motion}
The classical motion of a free  particle with mass $M$ 
on a conical surface may be determined from the Lagrangian, 
\bn
{\cal L}&=&\frac{1}{2}M  \left(dl/dt \right)^{2}
\nonumber
\\
&=&\alpha^{-2} \left[ \frac{1}{2} M \dot{{\bf x}}^{2} 
- \left(1-\alpha ^{2} \right)\frac{\ell^{2}}{2Mr^{2}} \right].
\label{lagc}
\en
The line element $dl$ is given by (\ref{dlc}), 
the dot denotes differentiation 
with respect to the time $t$,
${\bf x}:=\left( x^{1},x^{2} \right)$
and
${\ell}:={\bf x} \times M\dot{{\bf x}}$
is the kinematical angular momentum
(note that ${\bf a}\times {\bf b}:= \epsilon_{ij}a^{i}b^{j}$).
The Euler-Lagrange equations of motion, i.e. the geodesic equations
on the cone, follow from (\ref{lagc}),
\be
M\ddot{{\bf x}} 
= - \left( 1-\alpha^{2} \right) \frac{{\ell}^{2}}{Mr^{3}} 
    {\bf e}_{r},
\label{emc}
\ee
where ${\bf e}_{r}:={\bf x}/r$.
One can regard (\ref{emc}) as the equations of motion of a particle
moving on the Euclidean plane under the action of an angular-momentum 
dependent central force, which is attractive for
$\alpha <1$ (negative curvature) [see (\ref{ir})] and repulsive for 
$\alpha >1$ (positive curvature).
Clearly when $\ell =0$ the motion is radial
and uniform.

The geometrical force in (\ref{emc}) has  
the nature of an inertial force \cite{dow67,cle85}. 
In fact it ranks somewhere between
an inertial force and a Newtonian force.
Indeed by changing from 
$\left\{ x^{1}, x^{2}\right\}$ to
$\left\{X^{1}, X^{2}\right\} $, the force is ``gauged away" 
everywhere, apart from the borders of the wedge
which is consistent with the statements in the previous section.
Therefore, unless $\alpha =1$, trajectories crossing
these rays are broken  straight lines, with uniform motion
(dashed line in Fig. \ref{ocone}).
For $\alpha>1$ the particle disappears through the branch cut,
continuing into another Cartesian coordinate system.
Note that another way of seeing this is to consider (\ref{Boun}) 
and that the Lagrangian may be recast as
${\cal L}= M( \dot{ \rho}^{2} + 
\rho ^{2} \dot{\varphi }^{2})/2$,
which is the Lagrangian of a free particle
moving on the plane.

\begin{figure}[ht]
\centerline{\epsffile{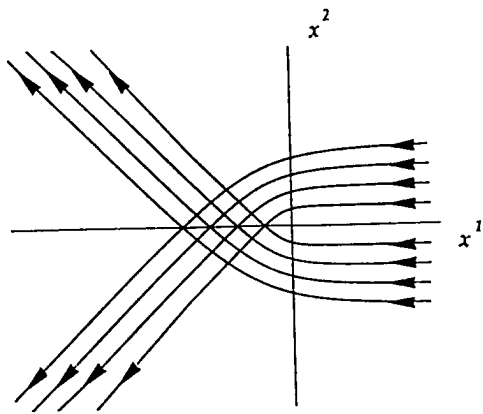}}
\caption{{\small Classical motion on a cone}.}
\label{p28}
\end{figure}

It is instructive to describe the classical motion in terms of 
polar coordinates $\left\{ r,\theta \right\}$.
Integration of (\ref{emc}) gives \cite{des88,cle85}
\bn
r^{2}\dot{\theta }=l&\hspace{1cm}&
\dot{r} ^{2}+\left(\frac{\alpha l}{r} \right)^{2}=\left(\alpha v 
\right)^{2},
\nonumber
\en
where {\it l} and {\it v} are constants corresponding to 
conservation of angular momentum and energy, respectively. 
A second integration gives
\bn
\left( vr\right) ^{2}=l^{2} 
+\left[\alpha v^{2}\left(t-t_{0} \right) \right] ^{2}
&\hspace{1cm}&l\tan\alpha\left(\theta-\theta_{0}\right)=
\alpha v^{2} \left(t-t_{0} \right),
\label{2i}
\en
resulting in the orbit equation
\be
vr\cos \alpha\left(\theta-\theta_{0} \right)=l.
\label{or}
\ee
For $l\neq0$ and choosing $\alpha \theta_{0}=\pi /2$, 
it follows from (\ref{2i}) and
(\ref{or}) that a particle initially at $r=\infty$ 
and $\theta_{i} = 0$, traveling say counterclockwise, winds 
\be
w=\left[1/2\alpha\right]
\label{wnumber}
\ee  
times
around the minimum radius $r_{0}=l/v$, and ends up at $r=\infty$
with 
$\theta_{f}=\pi(\alpha^{-1} -2w)$
($\left[ x \right] $ denotes the integral part of $x$). 
At $r_{0}$ 
the particle reaches a velocity
$v$ which is a maximum for $\alpha<1$ and a minimum for $\alpha>1$.
For $\alpha>1/2$ the winding number $w$ vanishes, 
which is obviously the case when the force in
(\ref{emc}) is repulsive ($\alpha>1$).
An unusual feature due to the localized 
curvature is that $\theta_{f}$ does
not depend either on the asymptotic velocity $\alpha v$ or on the impact
parameter $l/\alpha v$. Trajectories of particles with
different $v$ and $l$ are parallel to each other (in the sense that
the direction of the velocity depends on $\theta$ only) as shown
in Fig. \ref{p28}. Note that in terms of $\{X^{1},X^{2}\}$,
the particles travel parallel to the $X^{1}$ axis with 
velocity $-v$ before hitting the wedge
(see dashed line in Fig. \ref{ocone}).

A cone may be immersed in a magnetic field pointing
along its axis and homogeneous in that direction
by choosing a vector potential
${\bf A}({\bf x}, t)$ with vanishing component in
that direction.
The Lagrangian of a particle with charge $e$ moving in this 
background is given by
\be
{\cal L'}={\cal L} +\frac{e}{c} \dot{{\bf x}} \cdot {\bf A}.
\label{lagfc}
\ee
Thus the canonical momentum associated with ${\bf x}$ is
\bn
{\bf p} &=&\alpha^{-2} \left[ M\dot{{\bf x}} +\left(\alpha^{2} -1\right)
\frac{\ell}{r} {\bf e}_{\theta}  \right] +\frac{e}{c} {\bf A}
\nonumber\\
&=& M\dot{{\bf x}} +\left(\alpha^{-2} -1\right)
M\dot{r} {\bf e}_{r} +\frac{e}{c}{\bf A}.
\label{cm}
\en
It follows from (\ref{cm}) that
\be
{\bf x}\times {\bf p}=
{\ell} + \frac{e}{c}{\bf x} \times {\bf A}, 
\label{cam}
\ee
which is the usual expression for the canonical angular momentum
on the Euclidean plane.
The Hamiltonian is given by
\bn
{\cal H}&=&{\bf p}\cdot\dot{{\bf x}}-{\cal L}'\nonumber\\
&=&\frac{\alpha^{2}}{2M}\left( {\bf p}-\frac{e}{c}{\bf A}\right)^{2}
+\frac{\left(1- \alpha^{2} \right)}{2Mr^{2}} \left[{\bf x}\times
\left({\bf p} -\frac{e}{c}{\bf A}\right)\right]^{2}.
\label{ham}
\en

A magnetic flux $\Phi(t)$  running through the axis of the cone
can be introduced by choosing
\be
{\bf A} =\frac{\Phi}{2\pi r} {\bf e}_{ \theta}, 
\label{vp}
\ee
where ${\bf e}_{\theta}:=\left( -\sin \theta,\cos \theta \right)$. 
Now there is a new singularity at the origin which is 
a $\delta$-function magnetic field.
Due to the cylindrical symmetry, the canonical angular
momentum as given by (\ref{cam}) is conserved,
\be
\frac{d}{dt}\left(\ell + \frac{e\Phi}{2\pi c}\right)=0.
\label{ccam} 
\ee
The equations of motion following from 
(\ref{lagfc}) are given by (\ref{emc}),
with the induced electric force
$-e(\partial_{t}\Phi){\bf e}_{ \theta}/2\pi cr$
added on the r.h.s.. This electric force prevents
the orbital angular momentum $\ell$ being a
constant of motion, which is consistent with (\ref{ccam}).
Clearly the above expressions reduce to
the familiar ones on the plane \cite{jac83} when $\alpha=1$.

The Aharonov-Bohm set up may be combined with the 
conical geometry by taking $\Phi$ constant,
in which case the electric force 
vanishes, and (\ref{emc}) still holds. 
Obviously the same conclusion may be reached
by realizing that the electromagnetic Lagrangian
in (\ref{lagfc}) is a total derivative,
$e\Phi\dot{ \theta}/2\pi c$,
and consequently
does not affect classical motion. 
However, as is well known \cite{aha59}, 
this is not the case in quantum theory
which is the subject of the following sections.
(In the following sections, ${\bf A}$ is given by
(\ref{vp}) with constant $\Phi$.)

\section{Hamiltonian operator and 
stationary states}

The Hamiltonian operator $H$ can 
be obtained from (\ref{ham}) by the usual substitution \cite{des88},
${\bf p} \rightarrow -i\hbar{\bf \nabla}$.
The invariance of the
Schr\"{o}dinger equation under the gauge transformation  
${\bf A}({\bf x})\rightarrow  
{\bf A}({\bf x})+ {\bf \nabla}\chi ({\bf x})$ ,
$\psi({\bf x}, t) \rightarrow  
\exp \{i \left(e/\hbar c\right) \chi ({\bf x})\} \psi({\bf x}, t)$
allows the choice  
$\chi ({\bf x}) = (-\Phi /2\pi) \theta ({\bf x}) $, thereby 
gauging away the vector potential everywhere, except on an arbitrary ray 
where the polar angle
$\theta$ is a discontinuous function of ${\bf x}$
\cite{fel81}.
This singular gauge is analogous to the  singular 
Cartesian coordinates $\left\{ X^{1}, X^{2}\right\}$. 
It might have been anticipated that {\bf A} cannot 
vanish everywhere around the origin since
$\oint {\bf A}\cdot d{\bf x} = \Phi$. 
Similar reasoning may be applied
to the Levi-Civita connection of a conical geometry
(see \cite{bez89} and references therein).

Defining $\sigma:= -e\Phi/ch$, 
the transformed wave function is
\be
\psi'({\bf x}, t)= \exp\{i\sigma \theta\} \psi({\bf x}, t).
\label{nwf}
\ee
It then follows from (\ref{ham}), (\ref{npol}) and 
(\ref{Boun}) that
\bn
&&H= -\frac{\hbar^{2}}{2M}\frac{1}{\rho}\frac{\partial}{\partial
\rho}\left(\rho\frac{\partial}{\partial \rho}\right) +\frac{ L^{2}}
{2M\rho^{2}}
\label{hamo} \\
&&\psi(\rho,\varphi+2\pi\alpha)=\exp\{i2\pi\sigma\}\psi(\rho,
\varphi),
\label{boun}
\en
where 
\be
L:=-i\hbar \frac{\partial}{\partial \varphi}
\label{amo}
\ee
and the prime in $\psi '$ has been dropped. 
As $H$ in (\ref{hamo}) is just the free Hamiltonian operator on the plane
written in polar coordinates [which is not surprising since 
$\left\{ X^{1}, X^{2}\right\}$ is a (singular) Cartesian frame], 
the interaction with the magnetic flux and the conical
geometry manifest themselves only through (\ref{boun}).
In fact the twisted boundary condition (\ref{boun}) 
states that the wave function is not single valued,
for non integer values of the flux parameter $\sigma$,
and therefore is not continuous along some ray. This
corresponds to (and is compatible with) 
the fact that $H$, as given by 
(\ref{hamo}), disguises a singular vector potential
which, as mentioned above, is not defined everywhere.
Note also that according to (\ref{boun}) $\psi$
must either vanish or diverge at the origin
(for non integer $\sigma$), otherwise
an inconsistency results when a loop is shrunk around
$\rho=0$ \cite{dir36}.

Considering (\ref{hamo}) and (\ref{boun}), 
it follows from the Schr\"{o}dinger equation 
\be
\frac{d}{dt}
\int_{0}^{\infty}d\rho\ \rho \int_{0}^{2\pi\alpha}d\varphi\ 
\psi \psi ^{\ast}= 
\lim_{\rho \rightarrow 0}
\int_{0}^{2\pi\alpha}d\varphi\
\rho {\rm J}_{\rho},
\label{pc}
\ee
where ${\rm J}_{\rho}$ is the usual expression
for the radial component of the probability current
on the plane,
\bn
{\rm J}_{\rho}=
\frac{1}{M}{\rm Re}\left[\psi ^{\ast}
\frac{\hbar}{i}\frac{\partial\psi}{\partial \rho}
\right]
&&
{\rm J}_{\varphi}=
\frac{1}{M}{\rm Re}\left[\psi ^{\ast}
\frac{\hbar}{i\rho}\frac{\partial\psi}{\partial \varphi}
\right].
\label{rc}
\en
In obtaining (\ref{pc}), it has been assumed 
that the wave function vanishes at infinity.
The r.h.s. of (\ref{pc}) is the net probability crossing
an infinitesimally small circle around the singularity.
Equating (\ref{pc}) to zero amounts to the
statement that the singularity at the
origin is neither a source nor a sink (probability is conserved), 
which is automatically guaranteed if 
\be
\lim_{\rho \rightarrow 0}
\int_{0}^{2\pi\alpha}d\varphi\
\rho \left(\psi ^{\ast}\frac{\partial\phi}{\partial \rho}
-\frac{\partial\psi ^{\ast}}{\partial \rho}\phi \right)=0.
\label{gpc}
\ee
Expression (\ref{gpc}) is the condition for
self-adjointness of the Hamiltonian operator (\ref{hamo}),
$\langle\psi|H|\phi\rangle=
\langle\phi|H|\psi\rangle ^{\ast}$.

If $R_{k,m}(\rho)$ are functions such that
\be
\left[\rho\frac{\partial}{\partial
\rho}\left(\rho\frac{\partial}{\partial \rho}\right)
-\left(\frac{m+\sigma}{\alpha}\right)^{2}
+\left(k\rho\right)^{2}\right]R_{k,m}(\rho)=0,
\label{bde}
\ee
then 
\be
\psi_{k,m}(\rho,\varphi)
=\frac{1}{\sqrt{2\pi\alpha}}R_{k,m}(\rho)
e^{i(m+\sigma)\varphi/\alpha},
\label{eig}
\ee
where 
$0\leq k < \infty$ and  $m$ 
is an integer, are simultaneous eigenfunctions of
$ H$ and $L$ 
with eigenvalues 
$\hbar^{2}k^{2}/2M$ and $(m+\sigma)\hbar/\alpha$, 
respectively. 
Note that the effect of the magnetic flux
and of the conical geometry on the eigenvalues
of $L$ is to shift and to rescale them, 
respectively. For a particle
with spin there is also a shift due to a coupling
between the spin and the deficit angle
\cite{ger89a,ger90}.

In order that the stationary states $\psi_{k,m}$ 
span a space of wave functions in which probability
is conserved, they must satisfy (\ref{gpc}),
\be
\lim_{\rho \rightarrow 0}
\rho\left(R_{k',m}^{\ast}\frac{\partial R_{k,m}}{\partial \rho}
-\frac{\partial R_{k',m}^{\ast}}{\partial \rho}R_{k,m}\right)=0,
\label{gpce}
\ee
where the orthonormality relation
$\int_{0}^{2\pi\alpha}d\varphi\ \exp
\left\{ i\varphi (m-n)/\alpha\right \}=
2\pi\alpha\delta _{mn}$ 
has been used.
Expression (\ref{gpce}) is the condition for self-adjointness
of the square of the operator 
$P_{\rho}:= -i\hbar 
(\partial _{\rho}+ 1/2\rho)$,
and is itself self-adjoint if 
\be
\lim_{\rho \rightarrow 0}
\rho \left(\phi \psi ^{\ast} \right)=0,
\label{sap}
\ee
as may easily be shown by equating
$\int_{0}^{\infty}d\rho\ \rho
[\phi(P_{\rho}\psi)^{\ast}-
\psi^{\ast}(P_{\rho}\phi)]$ to zero.
One sees from (\ref{sap}) that $R_{k,m}(\rho)$
may diverge mildly at the origin [which also
applies to ${\rm J}_{\rho}$ in (\ref{pc})] and still be 
compatible with conservation of probability. A mild divergence
would not spoil the square integrability of the wave function,
since a behavior 
$\psi\sim 1/\rho^{\nu}$ with $\nu<1$ yields
$\int_{0}^{\epsilon}d\rho \rho|\psi|^{2}
\propto \epsilon^{2(1-\nu)}$ which vanishes with $\epsilon$.
Note that in order to satisfy (\ref{sap}) one must have $\nu<1/2$.
Obviously a logarithmic divergence also passes this test, since
it is weaker than the $1/\rho^{\nu}$ divergence.

In the following only wave functions
which are finite at the singularity will be considered. 
In quantum mechanics 
on the flux cone, finiteness is motivated by the fact that it 
arises naturally when the singularities are smoothed by some
regularization procedure \cite{kay91,bou92}.
[Divergent wave
functions have been considered in \cite{kay91,bou92}.]
Therefore one takes as solutions of (\ref{bde})
Bessel functions of the first kind
\be
R_{k,m}(\rho)=J_{|m+\sigma|/\alpha}(k\rho)
\label{bf}
\ee
which being finite at the origin satisfy (\ref{sap})
and consequently (\ref{gpce}). As a check we may 
verify that (\ref{bf}) indeed satisfies (\ref{gpce})
by observing that 
\be
\frac{dJ_{\nu}(x)}{dx}=\frac{1}{2}
\left[J_{\nu -1}(x)-J_{\nu +1}(x)\right].
\label{dbf}
\ee
Hence the complete set
of stationary states $\psi_{k,m}$ 
is given by (\ref{eig}) and (\ref{bf}).

A particular state to probe the
singularity can be found as follows.
The general expression for a stationary state
of energy $\hbar^{2} k^{2}/2M$ is given by
\be
\psi_{k}(\rho,\varphi)=
\sum_{m=-\infty}^{\infty} c_{m}\psi_{k,m}(\rho,\varphi).
\label{gst}
\ee
The coefficients $c_{m}$ are determined  by considering
the Fourier expansion of a plane wave 
\be
\exp\left\{-ia\cos\phi\right\}=
\sum_{m=-\infty}^{\infty}\left(-i \right)^{|m|}J_{|m|}(a)e^{im\phi},
\label{pw}
\ee
and letting the magnetic flux and the conical geometry
act on it. This involves shifting and rescaling the 
angular momentum of the modes, as mentioned previously,
leading to
\be
c_{m}^{(\delta ,\alpha)} :=\sqrt{2\pi\alpha}
\exp\left\{-\frac{i\pi}{2\alpha}|m-\delta|\right\},
\label{c}
\ee
where the flux parameter has been redefined to be 
$\delta := -\sigma$ in order to compare the result with earlier
work. The stationary state (\ref{gst}) with 
$c_{m}=c_{m}^{(\delta,\alpha)}$ will be denoted 
$\psi_{k}^{(\delta,\alpha)}$. Clearly
\bn
\psi_{k}^{(0,1)}(\rho,\varphi)&=&\exp\left\{-ik\rho\cos\varphi\right\}
\nonumber
\\
&\equiv &\exp\left\{-ikX^{1}\right\}
\label{pw0}
\en
(Note that when $\alpha=1$, both
$\{x^{1},x^{2}\}$ and $\{X^{1},X^{2}\}$ are genuine
Cartesian coordinates which coincide.).
The stationary states $\psi^{(\delta,1)}$ and $\psi^{(0,\alpha)}$
have previously been considered in the literature (in \cite{aha59}
and \cite{lan90, des88} respectively) in the context of 
scattering. The following section considers the probability 
fluid (see appendix) associated with $\psi_{k}^{(\delta,\alpha)}$.

\section{Quantum flow}
Consider the symmetries of  
the state $\psi^{(\delta,\alpha)}_{k}$.
By redefining the summation index in (\ref{gst})
it is straightforward to show that
\be
\psi^{(\delta+n,\alpha)}_{k}(\rho,\varphi)=
\psi^{(\delta,\alpha)}_{k}(\rho,\varphi),
\label{fs}
\ee
from which it follows  
that integer flux parameters are equivalent to zero. 
Also 
\be
\psi^{(-\delta,\alpha)}_{k}(\rho,\varphi)=
\psi^{(\delta,\alpha)}_{k}(\rho,-\varphi)
\label{ss}
\ee
implies that for vanishing flux parameter 
$\psi^{(0,\alpha)}_{k}(r,\theta)\equiv
\psi^{(0,\alpha)}_{k}(r,-\theta)$. Thus the 
quantum flow is symmetric with respect to
the $x^{1}$ axis and, consequently, no probability
from the upper half-plane passes to the lower half-plane
and vice versa. In other words the current lines
associated with the quantum flow must not cross the
$x^{1}$ axis, otherwise due to the symmetry, they
would intercept on this axis.
When the flux parameter is switched on
this symmetry is broken, implying that the flow corresponding to 
a charged particle is sensitive to the direction
(up or down) in which the magnetic flux runs.
By studying the symmetries (\ref{fs}) and (\ref{ss}),
one sees that 
\be
0\leq\delta\leq 1/2
\label{range}
\ee
covers all possible
behaviours of the flow and that $\delta=1/2$ also yields 
a flow symmetric with respect to the $x^{1}$ axis.
Expressions (\ref{fs}) and (\ref{ss}) generalize the 
known symmetries \cite{bel80,ola85} of the A-B set up 
to include the presence of a conical singularity.

Considering (\ref{gst}) and (\ref{c}) it follows
the probability density
\be
\psi^{(\delta,\alpha)}_{k}\psi^{(\delta,\alpha)^{\ast}}_{k}=
\sum_{m,m'=-\infty}^{\infty}
\exp\left\{i\Theta^{(\delta,\alpha)}_{m,m'}(\varphi)\right\}
J_{|m-\delta|/\alpha}(k\rho)J_{|m'-\delta|/\alpha}(k\rho),
\label{pd}
\ee
where
\be
\Theta^{(\delta,\alpha)}_{m,m'}(\varphi):=
\frac{1}{\alpha}
\left[\left(|m-\delta|-|m'-\delta|\right)\frac{\pi}{2}
-\left(m-m'\right)\varphi\right],
\label{tta}
\ee
satisfying
$\Theta^{(\delta,\alpha)}_{m,m'}=
-\Theta^{(\delta,\alpha)}_{m',m}$.
The probability current, 
when expressed with respect to $\left\{ X^{1}, X^{2}\right\}$,
has the familiar polar components on the Euclidean plane 
(\ref{rc}), from which it follows that 
\bn
{\rm J}_{\rho}^{(\delta,\alpha)}(\rho,\varphi)&=&\frac{\hbar k}{2M}
\sum_{m,m'=-\infty}^{\infty}
\sin\left\{\Theta^{(\delta,\alpha)}_{m,m'}(\varphi)\right\}
J_{|m-\delta|/\alpha}(k\rho)
\nonumber
\\
&&\times \left[J_{|m'-\delta|/\alpha-1}(k\rho) 
-J_{|m'-\delta|/\alpha+1}(k\rho)\right]
\label{rcur}
\en
and
\be
{\rm J}_{\varphi}^{(\delta,\alpha)}(\rho,\varphi)=\frac{\hbar}{M\rho}
\sum_{m,m'=-\infty}^{\infty}\frac{m'-\delta}{\alpha}
\cos\left\{\Theta^{(\delta,\alpha)}_{m,m'}(\varphi)\right\}
J_{|m-\delta|/\alpha}(k\rho)J_{|m'-\delta|/\alpha}(k\rho). 
\label{acur}
\ee
In deriving (\ref{rcur}), equality (\ref{dbf}) has been used.

In the following the quantum flow will be studied when 
\be
k\rho\ll 1
\label{reg1}
\ee
This amounts to consider the expansion
\be
J_{\nu}(z)=\left(\frac{z}{2}\right)^{\nu}
\left[\frac{1}{\Gamma(1+\nu)}-\frac{1}{\Gamma(2+\nu)}
\left(\frac{z}{2}\right)^{2}+ O(z^{4})\right],
\label{bee}
\ee
in (\ref{pd}), (\ref{rcur}) and (\ref{acur}),
keeping only the terms with small $m$ 
and $m'$. For simplicity the cone and the flux tube
will be considered separately.

It should be remarked that (\ref{reg1}) contrasts with the regime on  
scattering problems for which $k\rho\gg 1$.

\subsection{Conical singularity}
Using (\ref{tta}) and (\ref{bee}),
it follows from (\ref{pd}) that the first terms of the probability density
around a conical singularity are
\bn
\psi^{(0,\alpha)}_{k}\psi^{(0,\alpha)^{\ast}}_{k}&=&
1+\frac{\cos\{\pi/2\alpha\}\cos\{\varphi/\alpha\}}
{2^{1/\alpha-2}\Gamma(1+1/\alpha)}(k\rho)^{1/\alpha}
-\frac{1}{2}(k\rho)^{2}
\label{pdc}\\&&
+\frac{\alpha^{2}}{2^{2/\alpha-1}}
\left[\frac{1+\cos\{2\varphi/\alpha\}}
{\Gamma^{2}(1/\alpha)} +
\frac{\cos\{\pi/\alpha\}\cos\{2\varphi/\alpha\}}
{\alpha\Gamma(2/\alpha)}\right](k\rho)^{2/\alpha},
\nonumber
\en
where all terms of the order of $(k\rho)^{\lambda}$,
with $\lambda\leq 2$ and $1/2<\alpha<3/2$, have been considered. 
When $\alpha=1$ the corresponding (non zeroth) powers of $k\rho$
in (\ref{pdc}) cancel out as they should be since in the
absence of the conical singularity (and of magnetic flux) 
the probing state becomes a plane wave [see  (\ref{pw0})].

In studying the probability density (\ref{pdc}), one may be led
to conclude that the configuration space of a particle 
in the state $\psi^{(0,\alpha)}_{k}$
is simply connected, since 
$\psi^{(0,\alpha)}_{k}\psi^{(0,\alpha)^{\ast}}_{k}$
does not vanish at the conical singularity (i.e. in the
limit $k\rho\rightarrow 0$). However, the following considerations
may lead to a different interpretation.
A way of preventing a particle of getting
inside a disc centered at the origin is to 
impose boundary conditions such that 
the radial component of the probability current vanishes on the 
border of the disc
(and this may be implemented
without requiring that the wave function
itself vanishes at the border).
Thus no probability leaks into the hole and the 
current lines surround the disc without crossing it. This is
a non simply connected configuration space.
When the disc degenerates to a point at the origin, this
picture does not change. At the origin, which is now border of the disc, 
the radial probability current vanishes. 
It is reasonable to say that
the configuration space of this particle is $R^{2}-\{ 0\}$, although
the probability density may be non vanishing at the origin.
If this interpretation is adopted, it follows that before 
making any statement about  
connectivity of the configuration space of a particle on the cone, 
one should also study the probability current at the singularity.

Before determining the behaviour of the probability current at  
the conical singularity, consider its corresponding quantum potential
(\ref{defqpotential}). The term containing $(k\rho)^{1/\alpha}$ in (\ref{pdc})
satisfies the Laplace equation and, consequently, does not
give any information about the quantum potential (that is the reason
why higher order corrections in (\ref{pdc}) were considered).
For $1/2<\alpha<3/2$ it follows from (\ref{defqpotential}) and (\ref{pdc})
that the quantum potential
near the conical singularity is approximately given by
\be
V_{\alpha}(\rho)=-\frac{(\hbar k)^{2}}{2M}\frac{(k\rho/2)^{2/\alpha-2}}
{\Gamma^{2}[1/\alpha]},
\label{qpcone}
\ee
where an unimportant constant has been dropped. 
The fact that (\ref{qpcone}) is not a constant when the 
conical singularity is present ($\alpha\neq 1$) constitutes a
genuine quantum mechanical effect (a geometrical Aharonov-Bohm
effect). Far away from the conical singularity, for positive $X^{1}$, 
the state $\psi^{(0,\alpha<1/2)}$ behaves approximately 
as the plane wave (\ref{pw0}) \cite{lan90,des88}. Consequently,
in terms of  $\{X^{1},X^{2}\}$, the 
current lines of the quantum flow are approximately straight lines 
parallel to the $X^{1}$ axis, running to the left.
At this stage, the quantum flow coincides with a flow of classical
particles (see section 3). 
As the singularity is approached, the two flows depart from each other.
Near the conical singularity they differ radically - 
by differentiating the quantum potential (\ref{qpcone}),
it follows that the current lines are scattered away 
from the conical singularity
when $\alpha<1$ and bent towards it when $\alpha>1$,
whereas the classical particles experience no force 
(in the $\{X^{1},X^{2}\}$ frame).
When $\alpha=1$, the quantum potential (\ref{qpcone})
becomes constant and the classical and quantum flows coincide everywhere, 
as they must.

After this rather qualitative analysis, the
probability current will now be determined near the conical singularity.
By proceeding as one did to obtain (\ref{pdc}), from
(\ref{rcur}) and (\ref{acur}) it follows that
\be
{\rm J}_{\rho}^{(0,\alpha)}(\rho,\varphi)=-\frac{\hbar k}{M}
\left(\frac{k\rho}{2}\right)^{1/\alpha-1}
\left[\frac{\sin\{\pi/2\alpha\}\cos\{\varphi/\alpha\}}
{\Gamma(1/\alpha)}
+\frac{\sin\{\pi/\alpha\}\cos\{2\varphi/\alpha\}}
{2^{1/\alpha}\Gamma(2/\alpha)}(k\rho)^{1/\alpha}\right]
\label{radialc}
\ee
and
\be
{\rm J}_{\varphi}^{(0,\alpha)}(\rho,\varphi)=\frac{\hbar k}{M}
\left(\frac{k\rho}{2}\right)^{1/\alpha-1}
\left[\frac{\sin\{\pi/2\alpha\}\sin\{\varphi/\alpha\}}
{\Gamma(1/\alpha)}
+\frac{\sin\{\pi/\alpha\}\sin\{2\varphi/\alpha\}}
{2^{1/\alpha}\Gamma(2/\alpha)}(k\rho)^{1/\alpha}\right],
\label{angularc}
\ee
where terms
$O[(k\rho)^{2}]$ and $O[(k\rho)^{2/\alpha}]$
have been omitted for $\alpha\leq 1$ or $\alpha>1$, respectively. 
When $1/\alpha$ is even, one sees that the expressions (\ref{radialc})
and (\ref{angularc}) vanish, which is consistent with the fact 
that the stationary state $\psi^{(0,1/2n)}_{k}$ is a real function
\cite{des88}.
This effect becomes more intuitive by recalling that for even $1/\alpha$ 
classical particles are scattered backwards, and then
the scattered classical flow cancels the incident one, resulting
in a vanishing {\em net} classical flow.
Clearly when $\alpha=1$ 
(\ref{radialc}) and (\ref{angularc}) imply that 
${\rm J}_{\rho}^{(0,1)}(\rho,\varphi)=-(\hbar k/M)\cos\varphi$
and
${\rm J}_{\varphi}^{(0,1)}(\rho,\varphi)=(\hbar k/M)\sin\varphi$,
which are the polar components of the plane wave probability current, 
\be
{\bf{\rm J}}^{(0,1)}(\rho,\varphi)=-\frac{\hbar k}{M}{\bf e}_{1}
\label{ppcurr}
\ee
[${\bf e}_{1}:=(1,0)$] as expected.
From (\ref{radialc}) and (\ref{angularc}),
one sees that in the limit $\rho\rightarrow 0$ the
probability current either vanishes or diverges for $\alpha<1$
and  $\alpha>1$,
respectively. Note that although ${\rm J}_{\rho}^{(0,\alpha>1)}$
diverges as $\rho\rightarrow 0$, the r.h.s. of (\ref{pc}) vanishes,
which is not surprising since this was the criterion for choosing
the stationary states. The fact that for $\alpha<1$ the 
quantum flow avoids the conical singularity suggests that 
the corresponding configuration space is
non simply connected (and vice versa for $\alpha\geq 1$),
as discussed previously.
At this point it should be remarked that according to the authors of
\cite{bou92}, if $\psi\neq 0$ at the conical singularity,
the configuration space is simply connected and, consequently,
two identical particles on the plane 
($\alpha=1/2$) in the state $\psi$ ``collide''.
The analysis of the quantum flow above seems to suggest that this 
may be not the case.

Due to the presence of the wedge in the singular Cartesian coordinates
$\{X^{1},X^{2}\}$, for some purposes it is more convenient to describe the flow
in terms of the embedded Cartesian coordinates $\{x^{1},x^{2}\}$. By 
performing the coordinate transformation $X^{i}\rightarrow x^{i}$
it is straightforward to show that, in the   $\{x^{1},x^{2}\}$ frame,
probability current
is given by 
${\bf j}= {\rm j}_{r}{\bf e}_{r}+{\rm j}_{\theta}{\bf e}_{\theta}$,
with $ {\rm j}_{r}=\alpha {\rm J}_{\rho}$ and 
${\rm j}_{\theta}= {\rm J}_{\varphi}$.
(Note that the wave function and ${\bf j} $ satisfy the usual Cartesian
form of the continuity equation.)
Then, keeping only the leading contribution in (\ref{radialc}) 
and (\ref{angularc}), one finds
\be
{\bf j}^{(0,\alpha)}(r,\theta)=-\frac{\hbar k}{M}
\left(\frac{kr}{2\alpha}\right)^{1/\alpha-1}
\frac{\sin\{\pi/2\alpha\}}{\Gamma(1/\alpha)}
\left[{\bf e_{1}} +(\alpha-1)\cos\theta\ {\bf e}_{r}\right],
\label{pcsmall}
\ee
where the features mentioned above may easily be verified.
For example setting $\alpha=1$ in (\ref{pcsmall}) reproduces (\ref{ppcurr}).

\begin{figure}[ht]
\centerline{\epsffile{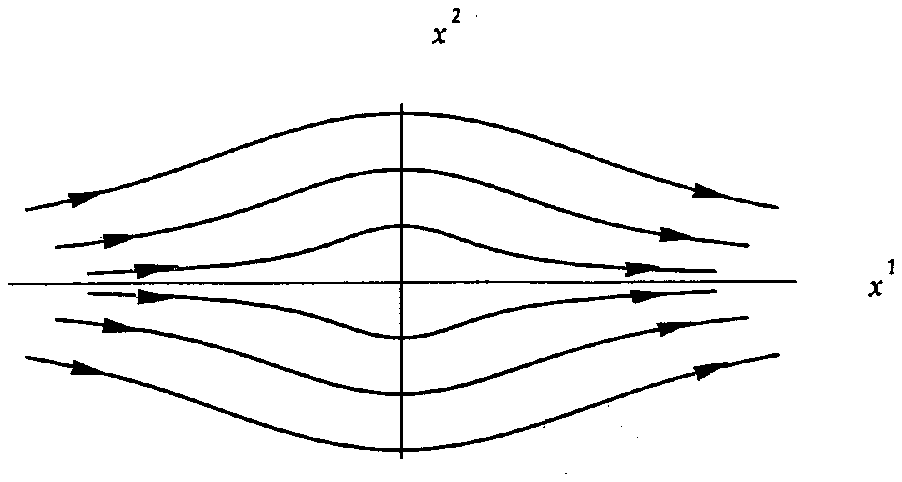}}
\caption{{\small Quantum flow when $\alpha<1$}.}
\label{p7}
\end{figure}

\begin{figure}[ht]
\centerline{\epsffile{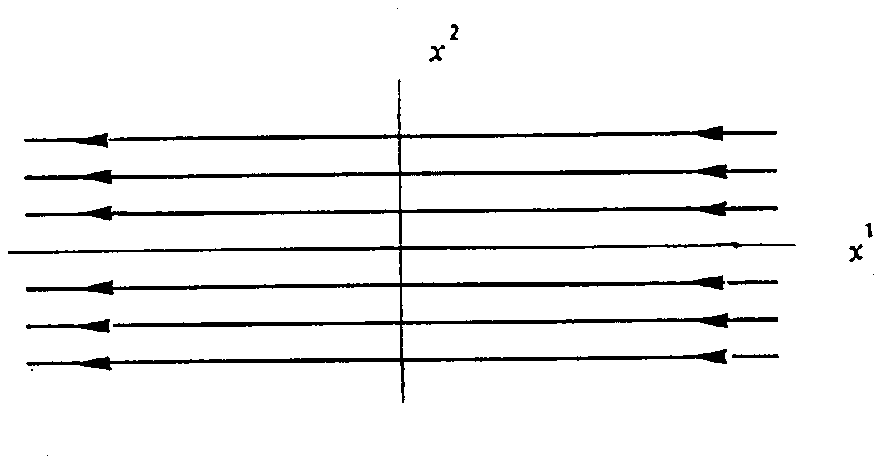}}
\caption{{\small Quantum flow when $\alpha=1$}.}
\label{p12}
\end{figure}

\begin{figure}[ht]
\centerline{\epsffile{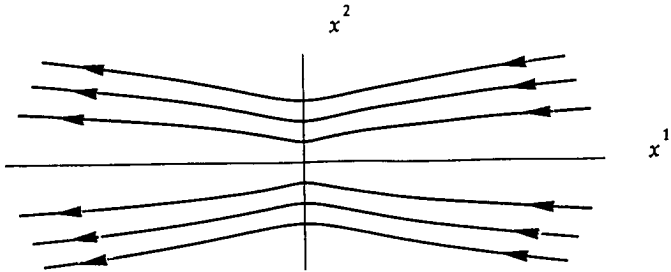}}
\caption{{\small Quantum flow when $\alpha>1$}.}
\label{p11}
\end{figure}

Figures \ref{p7}, \ref{p12} and \ref{p11}
show  the current lines associated with
${\bf j}^{(0,\alpha)}$ for $\alpha<1$, $\alpha=1$ and $\alpha>1$.
They have been obtained by plotting
the numerical integration of the equations resulting by equating
$d{\bf x}/d\lambda$ to the term between brackets in (\ref{pcsmall}),
\bn
\frac{dx^{1}}{d\lambda}=1+(\alpha-1)\frac{(x^{1})^{2}}{(x^{1})^{2}+
(x^{2})^{2}}&\hspace{1cm}&
\frac{dx^{2}}{d\lambda}=(\alpha-1)\frac{x^{1}x^{2}}{(x^{1})^{2}+
(x^{2})^{2}},
\nonumber
\en
where $\lambda$ is an arbitrary parameter.

At $\theta=0$, $\pm \pi/2$ and $\pi$ 
the flow runs parallel to the $x^{1}$ axis, as may be seen from
(\ref{pcsmall}). As the direction of ${\bf j}^{(0,\alpha)}$ in
(\ref{pcsmall}) for a given $\alpha$ is 
determined by $\theta$ only, the current lines
are parallel to each other (a feature shared with the classical flow).
However this is true only very close to the
singularity, where the subleading contributions in (\ref{radialc}) and
(\ref{angularc}) are not relevant. 
Apart from diverging at the conical singularity
for $\alpha>1$, the probability current is smooth
everywhere.

Consider more carefully the factor
$f(r,{\cal D}):=(kr)^{1/\alpha-1}$ in (\ref{pcsmall}).
For an infinitesimal $\epsilon>0$, $f(0,0+\epsilon)=0$,
$f(0,0)=1$  and  $f(0,0-\epsilon)=\infty$. This discontinuity
makes the behavior of the quantum flow on the Euclidean plane 
change abruptly in the presence of a tiny  deficit angle.
No such effect exists at classical level, where
the velocity of the particles varies smoothly with the deficit angle. 
This effect is less suprising by recalling that even a tiny
deficit angle $\epsilon$ corresponds to a delta function curvature 
which ``pierces'' the configuration space.

An unsuspected relationship between the winding number of classical orbits
and the quantum flow arises 
when studying the direction of the latter
[left or right, with the flow (\ref{ppcurr}) running to the left]. 
Such direction is determine by the 
factor $\sin\{\pi/2\alpha\}$ in (\ref{pcsmall}).
For $\alpha>1/2$ the quantum flow always runs to the left, and the classical
particles are scattered without winding around the conical singularity. For
$\alpha=1/2$ (classical backward scattering) the quantum flow stops 
and reverses its direction for $\alpha<1/2$, with the
classical particles winding once around the conical singularity.
It continues to run to the right until one decreases $\alpha$
to $1/4$ when another
classical backward scattering takes place with another interruption of the
flow. By further decreasing $\alpha$,
the quantum flow starts to run to the left, while
the classical particles wind twice around the conical singularity.
Generically, the direction of the quantum flow is controlled by the 
winding number $w$ defined in (\ref{wnumber}) -  the quantum flow runs
to the left for even $w$ and to the right for odd $w$.

From Fig. \ref{p7}, one sees that for $\alpha<1$ the quantum flow 
negotiates the 
conical singularity in a manner similar to the one in which a low
velocity fluid negotiates a cylinder. This analogy may be taken as
an evidence that the configuration space is not simply connected
when $\alpha<1$, as was suggested above. 
Note that  there are no vortices present
anywhere since the magnetic flux is switched off
[observe (\ref{observable})].

\subsection{Flux tube}
Now consider non-vanishing magnetic flux in the absence of a
conical singularity.
The analog of (\ref{pdc}) is
\bn
\psi^{(\delta,1)}_{k}\psi^{(\delta,1)^{\ast}}_{k}&=&
\frac{1}{2^{2\delta}\Gamma^{2}(1+\delta)}(k\rho)^{2\delta}
+\frac{\sin\{\varphi+\delta\pi\}}
{\Gamma(1+\delta)\Gamma(2-\delta)}k\rho
\label{pdfluxt}
\\
&&
-\frac{\sin \varphi}
{2^{2\delta}\Gamma(1+\delta)
\Gamma(2+\delta)}(k\rho)^{1+2\delta}
+\frac{1}{2^{2-2\delta}\Gamma^{2}(2-\delta)}(k\rho)^{2-2\delta},
\nonumber
\en 
where all terms of order $(k\rho)^{\lambda}$
were taken into account,
with $\lambda\leq 1$, and (\ref{range}) was considered.
Expression (\ref{pdfluxt}) agrees with the one in \cite{ola85}
where a slightly different method has been used.
For $\delta=0$ (\ref{pdfluxt}) reduces to unity up to a  $(k\rho)^{2}$
term, which would be canceled if higher powers  
of $k\rho$ in (\ref{pdfluxt})
had been kept. For non vanishing flux parameter,
the probability
density vanishes at the flux tube ($\rho=0$)
and the corresponding configuration space is non simply connected.
The expression for the quantum potential
corresponding to (\ref{pdfluxt}) will not be given here.
Instead a study of
the probability current itself is carried out in the following.

\begin{figure}[ht]
\centerline{\epsffile{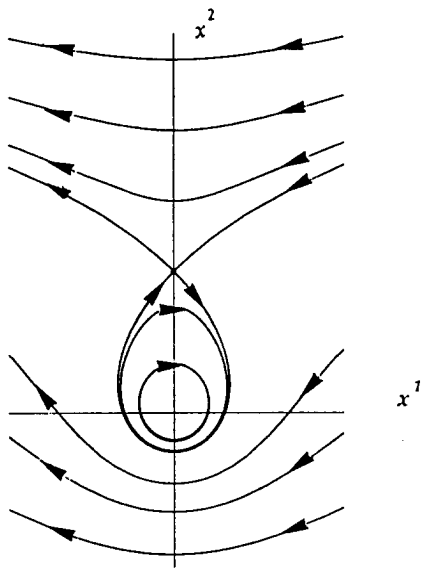}}
\caption{{\small Quantum flow around a flux tube}.}
\label{p27}
\end{figure}

As in the case of the conical singularity, 
(\ref{rcur}) and (\ref{acur}) may be used to obtain 
${\rm J}_{\rho}(\rho,\varphi)$ and 
${\rm J}_{\varphi}(\rho,\varphi)$. The expressions which correspond to
(\ref{radialc}) and (\ref{angularc}) have previously been found  
in \cite{ola85} (see also \cite{bel80}). 
The study will be limited to the case where the flux parameter 
is very small, viz. $\delta\ll 1$. In so doing,
${\bf J}= {\rm J}_{\rho}{\bf e}_{\rho}+{\rm J}_{\varphi}{\bf e}_{\varphi}$
reads approximately
\be
{\bf J}^{(\delta,1)}(\rho,\varphi)=\frac{\hbar k}{M}
\left[{\bf e}_{\delta}-\frac{\delta}{k\rho} {\bf e}_{\varphi}\right],
\label{pcsmallft}
\ee
where ${\bf e}_{\delta}$ is the unit vector ${\bf e}_{\rho}$
evaluated at $\varphi=\pi(1-\delta/2)$. When $\delta=0$ ,
(\ref{pcsmallft}) reduces to (\ref{ppcurr}), as should be. Clearly 
${\bf J}$ vanishes when 
${\bf e}_{\varphi} = {\bf e}_{\delta}$ and $k\rho=\delta$, so that
\be 
(\rho,\varphi)= (\delta /k,\ \pi(1-\delta)/2).
\label{bpoint}
\ee
It is also clear from (\ref{pcsmallft})
that the quantum flow 
(under the action of the corresponding quantum potential)
circulates around the origin  
when it is close to the origin \cite{bel80}. 
It turns out that (\ref{bpoint}) is a 
stagnation point and that there is a vortex around the flux tube,
which was expected by observing (\ref{observable}). 
In Fig. \ref{p27} the numerical solution of
\bn
\frac{dx^{1}}{d\lambda}=-1+(\delta/k)\frac{x^{2}}{(x^{1})^{2}+
(x^{2})^{2}}&\hspace{1cm}&
\frac{dx^{2}}{d\lambda}=-(\delta/k)\frac{x^{1}}{(x^{1})^{2}+
(x^{2})^{2}},
\nonumber
\en
is plotted, showing the main features of the quantum flow
near the flux tube. It is in agreement with \cite{ola85}
where the analytical expression for the current lines 
was given.

One sees from (\ref{pcsmallft}) that, as with the
conical singularity, the
delta function magnetic field at the origin
imparts a discontinuous change in the quantum flow 
of a charged particle.
Any tiny amount of magnetic flux changes the topology of the 
current lines (open lines become loops)
near the origin. From Fig. \ref{p27} one sees that,
as previously mentioned, the presence of the magnetic 
flux breaks the symmetry with respect the $X^{1}$ axis.
Notice that, as the distance from the origin increases 
(still for $k\rho\ll 1$), 
the flow gradually aproaches that of a plane wave
(\ref{ppcurr}), unlike the effect caused by the conical singularity.
An important distinction between the effects caused by the
conical singularity and the magnetic flux is that the latter only
affects charged particles ($\delta\neq 0$), whereas the former
affects all particles, indifferently. This fact may be seen as a
manifestation of the equivalence principle - geometrical
Aharonov-Bohm effects due to a conical singularity do not depend
on any particle attribute.

\section{Summary}
Summarizing, it was shown that the motion of a 
particle on a flux cone can be regarded
as a motion under the action of an angular-momentum 
dependent central force. 
Due to local flatness, quantization was implemented along usual
procedures in flat space.
By studying the probability fluid corresponding to a particular
stationary state (``plane wave'' on a flux cone), 
new effects were found.
For example it was shown that the winding number 
of classical orbits controls the direction
of the quantum flow on a cone. Classical flow (which is nearly trivial) and 
quantum flow were shown to depart from each other near the
singularity due to the presence of a non vanishing quantum force.
For the case of a conical singularity, the corresponding
quantum potential was determined and analyzed. 
The issue of connectivity of the configuration space
was treated and some interpretations were proposed.

Other boundary conditions at the singularity (those considered in
\cite{kay91,bou92}) may lead to new effects.
Another interesting extension of this work would be to consider  
relativistic particles. The use of this approach in the study
of quantum flow in the context of other geometries and topologies would 
be worthwhile, particularly in geometries with horizons.

\appendix

\section{Hydrodynamical approach to
quantum mechanics}
The analogies between quantum mechanics and fluid dynamics do not
stop at a continuity equation
which expresses local conservation of probability.				
The Schr\"{o}dinger equation can be rephrased as a set of 
hydrodynamical equations (see \cite{ola85} and references therein).
In order to derive them consider 
a particle with charge $e$ moving in the
Euclidean space (the generalization to arbitrary backgrounds is straightforward)
under the action of an electromagnetic field $({\bf{E}},{\bf{B}})$
with corresponding  potentials
$(\phi,{\bf A})$.
The Hamiltonian operator is then given by
\be
 H=\frac{1}{2M}\left(-i\hbar{\bf \nabla}-\frac{e}{c}{\bf A}\right)^{2}
+e\phi.
\label{emhamiltonian}
\ee
Now one rewrites the solution $\psi$ for the corresponding  
Schr\"{o}dinger equation as 
\bd
\psi(t,{\bf r})=
\varrho(t,{\bf r})e^{i\chi(t,{\bf r})}.
\ed
The real part of the Schr\"{o}dinger equation reads
\be
\hbar\frac{\partial\chi}{\partial t} +\frac{\hbar^{2}}{2M}
\left[\nabla\chi-\frac{e}{c\hbar}{\bf A}\right]^{2}+e\phi+V=0,
\label{hjequation}
\ee
where
\bn
V(t,{\bf r})&:=&-\frac{\hbar^{2}}{2M}\frac{\nabla^{2}\varrho}{\varrho}
\nonumber
\\
&\equiv&-\frac{\hbar^{2}}{2M}\frac{\nabla^{2}(\psi\psi^{\ast})^{1/2}}
{(\psi\psi^{\ast})^{1/2}}.
\label{defqpotential}
\en
The imaginary part reads
\be
\frac{\partial\varrho^{2}}{\partial t}+\nabla\cdot(\varrho^{2}{\bf v})=0
\label{cequation}
\ee
with 
\be
{\bf v}:=[\hbar\nabla\chi-e{\bf A}/c]/M,
\label{fvelocity}
\ee
leading to the constraint
\be
\nabla\times{\bf v}=-\frac{e}{Mc}{\bf B}.
\label{constraint}
\ee
Now one sees that 
$\varrho^{2}{\bf v}$ is the probability current 
and therefore (\ref{cequation}) is just
the continuity equation. 
Note that by integrating (\ref{constraint}) along a closed loop it follows
\be
\oint{\bf v}\cdot d{\bf l}=-\frac{e\Phi}{Mc},
\label{observable}
\ee
where $\Phi$ is the magnetic flux enclosed by the loop.

The last step is to apply $\nabla$ on (\ref{hjequation}), after which one 
is left with a set of equations
of motion for a fluid of density $\varrho^{2}$ and velocity ${\bf v}$,
\bn
M\frac{d}{dt}{\bf v}(t,{\bf r})&\equiv&
M\left[\frac{\partial{\bf v}}{\partial t}+({\bf v}\cdot\nabla){\bf v}\right]
\nonumber\\
&=&e{\bf E}+\frac{e}{c}{\bf v}\times{\bf B}-\nabla V.
\label{hdequation}
\en
Thus, the wave like Schr\"{o}dinger equation
[where the function to be determined is $\psi$ for a 
given configuration of potentials $(\phi,{\bf A})$] 
has been replaced by
the hydrodynamical equations (\ref{cequation}) and (\ref{hdequation}) 
with the constraint (\ref{constraint}) [where the functions to be determined
are $(\varrho,{\bf v})$ for a given configuration of electromagnetic field
$({\bf{E}},{\bf{B}})$].
These equations govern the behaviour of the quantum flow.

Whereas the Schr\"{o}dinger equation is more appropriate
to study wave like features of quantum mechanics, the hydrodynamical 
equations are more appropriate to study  particle like features.
To see this, assume that the quantum force $-\nabla V$ in  
(\ref{hdequation}) is negligible when compared with the Lorentz force
(and other forces that one might have considered). In this ``classical limit'',
(\ref{hdequation}) reduces to the equations of motion for a
flow of non-interacting classical particles.
Note that (\ref{hjequation}) is the corresponding 
Hamilton-Jacobi equation where the action
is identified with $\hbar\chi$. It is clear that the quantum potential $V$
represents the departure from the classical motion. In regions where the 
quantum potential is relevant, the classical and quantum flow may differ
considerably from each other.
For example, in a region of vanishing field strengths
the motion of the classical flow is trivial,
since the Lorentz force vanishes there. The motion of the 
quantum flow, on the other hand,
may be quite elaborate due to the presence of the 
quantum force $-\nabla V$ in (\ref{hdequation}). 
A non vanishing quantum force is the essence of  
Aharonov-Bohm like effects.

\vspace{5 mm}
{\bf Acknowledgements}.
The author is grateful to George Matsas for reviewing
the manuscript and for clarifying discussions.

\end{document}